# Unveiling the quantum critical point of an Ising chain

Y. F. Dai, H. Zhang, S. Y. Zhou, B. Y. Pan, X. Qiu, X. C. Hong, T. Y. Guan, J. K. Dong, Y. Chen, & S. Y. Li[*]

*Department of Physics, State Key Laboratory of Surface Physics, and Laboratory of Advanced Materials, Fudan University, Shanghai 200433, China*

**Quantum phase transitions occur at zero temperature upon variation of some nonthermal control parameters. The Ising chain in a transverse field is probably the most-studied model undergoing such a transition, from ferromagnetic to paramagnetic state[1,2]. This model can be exactly solved by using a Jordan-Wigner transformation, which transforms the spins into noninteracting spinless fermions[1]. At the quantum critical point, the magnetic excitations can carry arbitrarily low energy and dominate the low temperature properties. Here we report the unveiling of such quantum critical point in quasi-one-dimensional Ising ferromagnet $CoNb_2O_6$ by ultra-low-temperature thermal conductivity measurements. We find that in the paramagnetic state, phonons are scattered by the magnetic excitations above certain temperature $T_s$, which corresponds to a gap. As predicted by the theoretical model[1], this gap linearly goes to zero with decreasing the transverse field, thus determining the quantum critical point of the Ising chain.**

During last two decades, quantum phase transitions (QPTs) have attracted great attentions in condensed matter physics[1-4]. In contrast to a thermal phase transition tuned by temperature, a QPT is tuned at zero temperature by nonthermal control parameter such as chemical doping, pressure, and magnetic field. The quantum fluctuations near the quantum critical point (QCP) may involve many important issues, for example, high-temperature superconductors[5,6], heavy-fermion systems[7], and Bose-Einstein condensation in magnetic insulators[8].

While the underlying physics of the QPTs are complex, there are some relatively simple models which can display the basic phenomena of QPT[1]. Among them, the Ising chain in a transverse field (TFIC) is probably the most-studied model. It is described by the Hamiltonian[1]

$$H = -J\sum_i (\hat{\sigma}_i^z \hat{\sigma}_{i+1}^z + g\hat{\sigma}_i^x) \tag{1}$$

where $J > 0$ is the coupling between the spins $\hat{\sigma}^z$ on site $i$ and $i+1$, which prefers a ferromagnetic state. The $Jg$ in the second term represents the external transverse field which disrupts the magnetic order. When $g < 1$, the system is in a ferromagnetic ordered state. When $g$ is increased above the QCP $g_c = 1$, the system undergoes a phase transition into a paramagnetic state. By using the Jordan-Wigner transformation, the spins can be transformed to noninteracting spinless fermions, and the minimum single-particle excitation energy, or the energy gap is[1]

$$\Delta = 2J|1-g|. \tag{2}$$

This gap vanishes at $g_c = 1$.

Despite the simplicity, experimental realization of the TFIC model is very rare and the gap behaviour of equation (2) has not been observed so far. One difficulty is to find an Ising chain compound with sufficient low $J$, so that the QCP can be reached by a standard laboratory magnet (10 T ~ 1 meV). Another difficulty is that in realistic materials, there exist inevitable interchain couplings. This may bring in more complex physics into the system[9,10].

Only very recently, it was shown that the insulating quasi-one-dimensional (1D) Ising chain compound $CoNb_2O_6$ may be an ideal system to investigate the quantum criticality of the TFIC model[11]. In $CoNb_2O_6$, the magnetic $Co^{2+}$ ions are arranged into zigzag chains along the orthorhombic $c$ axis. While the dominant interaction is the intrachain ferromagnetic coupling $J \sim 1.94$ meV, there also exist weak interchain antiferromagnetic coupling $J_1$, $J_2 \ll J$ (ref. 11, 12). To the mean-field approximation, the weak interchain couplings can be treated as an effective longitudinal field[10]. Neutron scattering experiments have demonstrated an emergent $E_8$ symmetry of the spin excitations just below the critical field $H_c$ (ref. 11), which has been predicted by Zamolodchikov two decades ago for TFIC model adding a finite longitudinal field[9]. However, limited by the wave vector they were choosing for the inelastic scattering experiments, they can not observe the complete softening of the gap, and a gap about 0.36 meV (~ 4 K) remains at $H_c \approx 5.5$ T (ref. 11). Therefore it will be very important to find out whether the true gap goes to zero or not at the critical field with other experimental techniques. In fact, previously for the three-dimensional (3D) Ising ferromagnet $LiHoF_4$, the gap of the magnetic excitations never goes to zero at the critical field, due to the hyperfine coupling between the electronic and nuclear moments[13].

Ultra-low-temperature thermal conductivity measurement is a useful tool to probe the low-energy magnetic excitations[14,15]. Unlike the neutron scattering experiment, it is sensitive to all wave vectors, thus should be able to detect the true gap behaviour near the critical field of $CoNb_2O_6$. Here we present the thermal conductivity measurements of $CoNb_2O_6$ single crystal in transverse magnetic field $H \parallel b$. We find that at the critical field $H_c$, magnetic excitations strongly scatter phonons down to the lowest temperature of our experiment (60 mK ~ 5 μeV), therefore they are essentially gapless. In the paramagnetic state above $H_c$, these magnetic excitations develop a gap with linear field dependence, as expected from the TFIC model.

Fig. 1a shows the magnetization of $CoNb_2O_6$ single crystal in a field $H$ = 5 Oe, applied along *a*-, *b*-, and *c*-axis. For $H \parallel a$, there is an incommensurate spin-density-wave transition at $T_{N1}$ = 2.95 K and a commensurate antiferromagnetic transition at $T_{N2}$ = 1.97 K (ref. 16, 17), indicating by the arrows. These 3D transitions result from the interchain couplings. The *T*-independent and small magnetization in $H \parallel b$ shows that the *b*–axis is perpendicular to the local easy-axis lying in the *ac* plane[17]. The behaviours of these magnetizations are consistent with previous reports[16,18].

Fig. 1b shows the thermal conductivity κ/*T* versus $T^{1.76}$ for $CoNb_2O_6$ single crystal along the *c*–axis at zero field and $H$ = 4.7 T, respectively. At zero field, the data below 0.3 K can be fitted by κ/*T* = $a + bT^\alpha$ (ref. 19), with $a$ = -0.005 ± 0.007 mW $K^{-2}$ $cm^{-1}$ and α = 1.76 ± 0.03. This is the typical phonon conductivity behaviour of an insulator or *s*-wave superconductor, for phonons in the boundary scattering limit with specular reflections[19]. Applying a field $H$ = 4.7 T along the *c*-axis does not change κ/*T*. The reason is apparent from its magnetic structure in longitudinal field $H_{\parallel c}$ (ref. 17, 20). At zero field and below $T_{N2}$ = 1.97 K, the moments of $Co^{2+}$ ions point along four directions, contained in the two equivalent local easy-axis in the *ac* plane with an angle γ ≈ ±30° with respect to the *c*-axis[17,20]. Magnetization and neutron scattering experiments have shown that longitudinal field $H_{\parallel c}$ only causes two spin-flip transitions at fields lower than 0.4 T (refs. 17, 20). In $H_{\parallel c}$ = 4.7 T, the moments will point along only two directions now, but still contained in the two local easy-axis. From the aspect of quasiparticle heat transport, such an ordered state is no different from the zero-field state, therefore one does not expect any change of the thermal conductivity κ/*T*. When a transverse field $H$ = 4.7 T is applied along the $b$-axis, however, a strong suppression of $\kappa/T$ is observed.

Fig. 2 presents the evolution of $\kappa/T$ in transverse field $H_{\parallel b}$ up to 7 T. Let us focus on the high field data first. In Fig. 2b from 5 to 7 T, the $\kappa/T$ curves overlap with the zero-field curve below certain temperatures $T_s$. To precisely determine $T_s$, the difference between $\kappa(H)$ and $\kappa(0T)$ are plotted in Fig. 2c. The temperature at which $\Delta\kappa$ starts to deviate from the dashed horizontal line $\Delta\kappa = 0$ is defined as $T_s$, indicated by the arrows. This result strongly suggests that there exist some excitations to scatter the phonons above $T_s$. For each field, $T_s$ corresponds to the gap of these excitations. In Fig. 3b, the gap ($T_s$ in Fig. 2c) is plotted versus magnetic field. It shows a nice linear dependence, exactly as the gap behaviour of the magnetic excitations in the paramagnetic state of the TFIC model[1], despite the existence of weak interchain couplings in $CoNb_2O_6$. The physical reason may be that in the disordered phase, all moments point along the transverse field direction ($b$-axis for $CoNb_2O_6$), thus the effective interchain coupling is zero on average. Hence the system behaves like a pure 1D system[1]. We note that the energy of the spin-flip quasiparticle in the paramagnetic state measured by neutron scattering experiments also shows roughly linear field dependence, when slightly away from the critical field[11].

In Fig. 3b, a linear fit of the gap gives the quantum critical point $H_c$ = 4.5 T, where the gap is zero. At the QCP, phonons are scattered down to the lowest temperature of our experiments (60 mK ~ 5 μeV), as shown in Fig. 2a. We do not observe significant positive contribution of these magnetic excitations to the thermal conductivity. This may result from their very low velocity which is proportional to the intrachain coupling $J$, according to $\kappa = \frac{1}{3}Cvl$, with $C$ the specific heat, $v$ the velocity, and $l$ the mean-free-path.

Next we turn to the ordered state below $H_c$. The interplay of quantum criticality and geometric frustration has made the ordered state of $CoNb_2O_6$ very complex[12]. Considering the frustration in the triangular lattice perpendicular to the chain direction (for example, $J_2/J_1 = 0.9$, the Fig. 3c in ref. 12), the phase diagram at zero temperature is very rich with four different states upon increasing transverse field: Néel, ferrimagnetic (FR), incommensurate spin-density wave (IC), and paramagnetic phases[12]. There is also a region of homogeneous coexistence the FR and the IC states[12]. While the neutron scattering experiment only shows a magnetic phase transition at $H_c \approx 5.5$ T (ref. 11), the full phase diagram of $CoNb_2O_6$ needs to be experimentally established to compare with the theoretical one[12].

In Fig. 2a, the thermal conductivity $\kappa/T$ in the ordered state $H_{\|b} = 3$ and 4 T are very different from that in the paramagnetic state $H_{\|b} \geq 5$ T. The $\kappa/T$ in these two fields are suppressed below the zero-field curve, in the whole temperature range from 60 mK to 0.8 K. These results show that there exist gapless excitations to scatter phonons in the ordered state $H_{\|b} = 3$ and 4 T. While we are not sure whether these gapless excitations are some 3D magnetic fluctuations in the complex ordered state, their existence has prevented us to probe the gap behaviour of the 1D magnetic excitations below $H_c$. The 1D domain-wall quasiparticals with finite gap have been clearly observed in the ordered state by neutron scattering experiments[11].

For comparison, the field dependence of the lowest-energy mode measured by neutron scattering experiments[11] is reproduced in Fig. 3a. From Fig. 3a and 3b, there are two major differences between the two $\Delta(H)$. First, our gap goes to zero at QCP, while a gap about 0.36 meV (~ 4 K) remains at the critical field $H_c$ in ref. 11. The incomplete softening of the gap has been attributed to the scattering plane (3.6, 0, 0) they were using to do the inelastic scattering experiments. As described in ref. 11, a complete gap softening is only expected at the location of the 3D magnetic long-range order Bragg peaks (0, 0.34, 0). Secondly, our QCP ($H_c$ = 4.5 T) is lower than the critical field ($H_c$ = 5.5 T) in ref. 11. The reason may relate to the details of different experimental technique. Some other measurements, such as magnetization and heat capacity in transverse field and at ultra-low-temperature may resolve this discrepancy.

To conclude, we have detected the true gap of the magnetic excitation in the paramagnetic state of the quasi-1D Ising ferromagnet $CoNb_2O_6$, using ultra-low-temperature thermal conductivity measurements. This gap shows a linear field-dependence, and goes to zero at the QCP. Our results have experimentally confirmed the exact solution of the Ising chain in a transverse magnetic field, one of the most-studied models in condensed matter physics.

**METHODS SUMMARY**

Single crystals of $CoNb_2O_6$ were grown by the float-zone method, as in ref. 18. Small samples were cut along the crystallographic *a*-, *b*-, and *c*-axis. The dc magnetizations of a sample with mass 11.3 mg were measured by a SQUID (Quantum Design) in a field of 5 Oe, applied parallel to the *a*-, *b*-, and *c*-axis. For heat transport measurements, the samples were polished to a bar shape of typical dimensions 2.0 × 0.5 × 0.5 mm$^3$, with the longest dimension along the chain direction (*c*-axis). Four contacts were made by silver epoxy. Thermal conductivity along *c*-axis were measured in a dilution refrigerator, using a standard four-wire steady-state method with two $RuO_2$ chip thermometers, calibrated in situ against a reference $RuO_2$ thermometer. Magnetic fields were applied along the *b*- and *c*-axis, respectively. To avoid any movement of the sample in high fields, one end of the sample was firmly glued to a copper block with silver epoxy.

**Acknowledgements** We thank S. L. Li for discussions. This work is supported by the Natural Science Foundation of China, the Ministry of Science and Technology of China (National Basic Research Program No: 2009CB929203), Program for New Century Excellent Talents in University, Program for Professor of Special Appointment (Eastern Scholar) at Shanghai Institutions of Higher Learning, and STCSM of China (No: 08dj1400200 and 08PJ1402100).

**Author Information** Reprints and permissions information is available at www.nature.com/reprints. The authors declare no competing financial interests. Correspondence and requests for materials should be addressed to S. Y. Li (shiyan_li@fudan.edu.cn).

**Figure 1 | Magnetization and thermal conductivity of $CoNb_2O_6$ single crystal.**

**a,** Magnetization of $CoNb_2O_6$ single crystal in $H$ = 5 Oe, applied along $a$-, $b$-, and $c$-axis. The two 3D magnetic transitions, resulting from weak interchain couplings, are indicated by arrows. **b,** $\kappa/T$ vs $T^{1.76}$ in zero field and $H$ = 4.7 T, respectively. The solid line is a fit of the zero-field data to $\kappa/T = a + bT^{\alpha}$ between 60 mK and 0.3 K, with $a$ = -0.005 $\pm$ 0.007 mW / K$^2$ cm and $\alpha$ = 1.76 $\pm$ 0.03. This is the typical phonon conductivity behaviour of an insulator. While a longitudinal field $H_{\parallel c}$ = 4.7 T does not change $\kappa/T$, a transverse field $H_{\parallel b}$ = 4.7 T strongly suppresses $\kappa/T$.

**Figure 2 | Evolution of the thermal conductivity in transverse fields.**

**a,** $\kappa/T$ vs $T^{1.76}$ in $H_{\parallel b}$ = 0, 3, 4, and 4.5 T. The suppression of $\kappa/T$ in 3 and 4 T down to 60 mK suggests that there exist some gapless excitations to scatter phonons. **b,** $\kappa/T$ vs $T^{1.76}$ in $H_{\parallel b}$ = 0, 5, 5.5, 6, 6.5, and 7 T. **c,** $\kappa(H)$ - $\kappa(0T)$ vs $T$ for the data in **b**. The arrows denote the temperatures $T_s$, above which $\kappa(H)$ - $\kappa(0T)$ starts to deviate from the dashed horizontal line $\Delta\kappa$ = 0. Above $T_s$, magnetic excitations start to scatter phonons, therefore $T_s$ corresponds to the gap of these magnetic excitations.

**Figure 3 | Field dependence of the gap.**

**a,** Softening of the energy gap for domain-wall and spin-flip quasiparticles measured by neutron scattering experiments[11]. The gap does not close at the critical field $H_c \approx 5.5$ T, since the measurements were in a scattering plane (3.6, 0, 0) where no Bragg peaks occur. A complete gap softening is only expected at the location of the 3D magnetic long-range order Bragg peaks (0, 0.34, 0)[11]. **b,** Field dependence of the gap of the magnetic excitations in the paramagnetic state, measured by heat transport in this work. Note that the vertical scales in **a** and **b** are different by one order (1 meV = 11.6 K). The solid line is a linear fit to the data, which extrapolates to $H_c \approx 4.5$ T, the QCP. The dash line is the gap in the ordered state, expected for pure Ising chains. However, some gapless excitations in the complex ordered state of $CoNb_2O_6$ have prevented us to probe this gap with our heat transport measurements (see text).

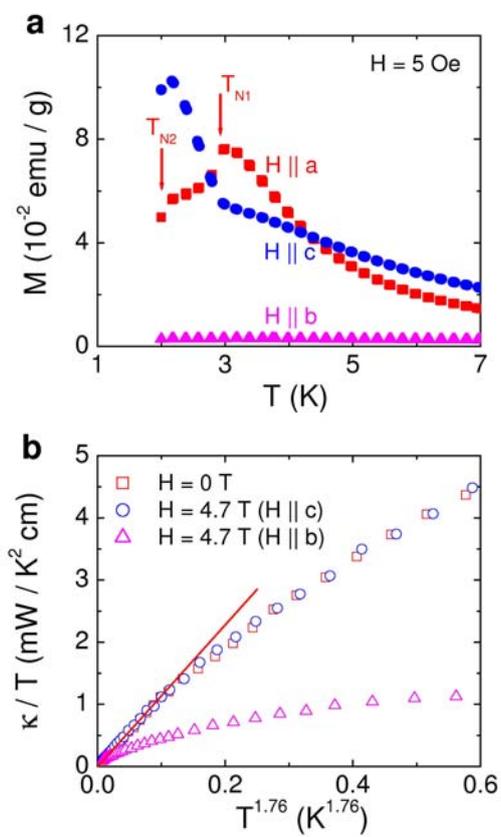

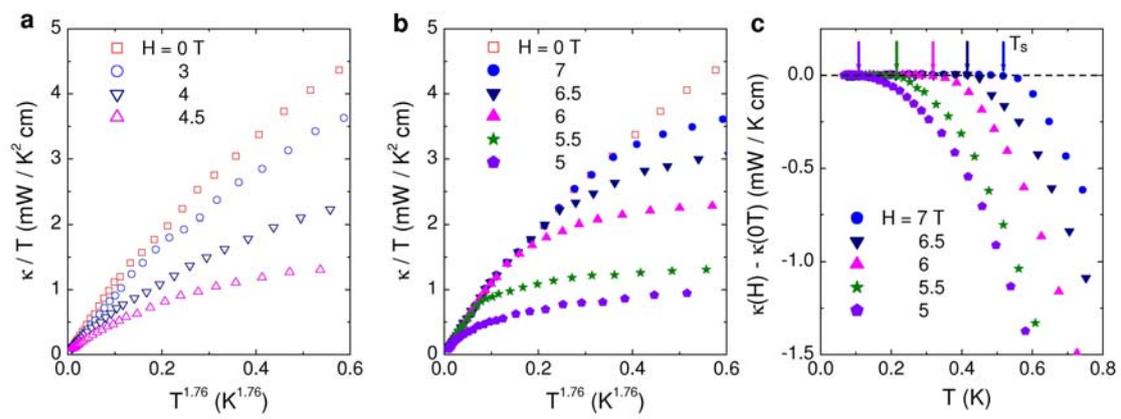

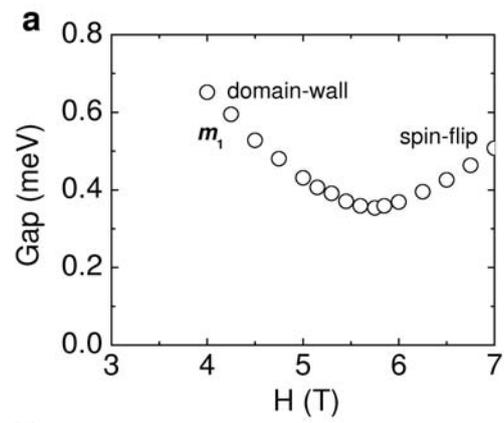

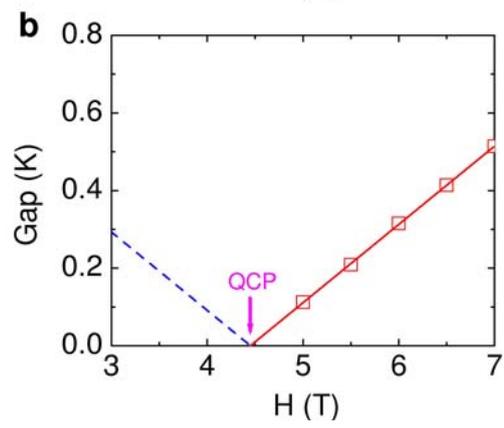